\documentclass[12pt]{article}
\usepackage{a4}
\usepackage{amssymb}
\usepackage{cite}
\newcommand{\be}{\begin{equation}}
\newcommand{\ee}{\end{equation}}
\newcommand{\bea}{\begin{eqnarray}}
\newcommand{\eea}{\end{eqnarray}}

\usepackage{hyperref}
\def\stackreb#1#2{\ \mathrel{\mathop{#1}\limits_{#2}}}

\begin{document}
\thispagestyle{empty}
\def\thefootnote{\fnsymbol{footnote}}
\begin{center}\Large
From rarefied elliptic beta integral to \\
parafermionic star-triangle relation  \\
\end{center}

\vskip 0.2cm

\begin{center}
Gor Sarkissian$^{1,3}$
\footnote{sarkissn@theor.jinr.ru, gor.sarkissian@ysu.am}
 and Vyacheslav P. Spiridonov $^{1,2}$\footnote{spiridon@theor.jinr.ru}
\end{center}
\begin{center}
$^1$ Bogoliubov Laboratory of Theoretical Physics, JINR,\\
Dubna, Moscow region, 141980 Russia\\
\end{center}
\begin{center}
$^2$ St. Petersburg Department of the Steklov Mathematical Institute
of Russian Academy of Sciences, St. Petersburg, 191023 Russia\\
\end{center}
\begin{center}
$^3$ Department of Physics, \ Yerevan State University,\\
Alex Manoogian 1, 0025\, Yerevan, Armenia
\end{center}
\vskip 1.5em
\begin{abstract} \noindent
We consider the rarefied elliptic beta integral in various limiting forms.
In particular, we obtain an integral identity for parafermionic
hyperbolic gamma functions which describes the star-triangle
relation for parafermionic Liouville theory.
\end{abstract}

\newpage
\tableofcontents
\newpage
\section{Introduction}
Full understanding of the properties of non-rational $2d$ conformal field theories is one of the most
important questions in string theory, quantum field theory and mathematical physics.
The most important well known examples of the non-rational conformal field theory are Liouville field theory (LFT)
and its various generalizations, like supersymmetric extensions, parafermionic extensions, Toda field theory, etc.
Already in the seminal works \cite{Dorn:1994xn,Zamolodchikov:1995aa}, where the three-point function in LFT was constructed,
an important role of some special function $\Upsilon_b$ emerged. Studies of the
fusion matrix \cite{Ponsot:1999uf} and boundary correlation functions
\cite{Fateev:2000ik,Ponsot:2001ng} required the use of another related
function -- the noncompact quantum dilogarithm $S_b$ \cite{F95},
which is called also the hyperbolic gamma function \cite{ruij}
(we follow the latter terminology).
Both of these functions are constructed
out of the Barnes double gamma function $\Gamma_b$.
Study of $N=1$ supersymmetric LFT showed that description of the three-point functions \cite{Poghosian:1996dw,Rashkov:1996np}, boundary correlation functions \cite{Fukuda:2002bv} and
fusion matrix \cite{Hadasz:2007wi,Chorazkiewicz:2008es} requires the use of supersymmetric generalizations
of these functions: $\Upsilon_i$, $\Gamma_i$ and $S_i$, where $i=0,1$.
In \cite{Bershtein:2010wz}, three-point functions were studied in parafermionic LFT, which is LFT
coupled with $\mathbb{Z}_N$ parafermions. It was shown there that three-point functions can be written using parafermionic generalizations of $\Upsilon_b$: $\Upsilon_k$, where $k=0,\ldots, N-1$.

Analysis of the bootstrap relation and boundary three point function in the LFT carried out in \cite{Sarkissian:2011tr} demonstrated that
some fundamental relations between fusion matrix and three-point functions established in rational conformal
field theories hold also in LFT. On the other hand, it was shown in
\cite{Vartanov:2013ima} that the expressions in the above mentioned
works indeed satisfy these relations due to a particular star-triangle
relation \cite{kashaev,volkov} for the hyperbolic gamma functions,
which corresponds to the Faddeev-Volkov model \cite{Bazhanov:2007vg}
(a more complicted star-triangle relation leading to a generalization
of the latter model was considered in \cite{Spiridonov:2010em}).
Similar analysis of the  Neveu-Schwarz sector of $N=1$ supersymmetric LFT
\cite{Poghosyan:2016kvd} showed that the corresponding relations
are implied by the generalization of considerations of \cite{kashaev}
to supersymmetric hyperbolic gamma functions found in \cite{Hadasz:2013bwa}.

This state of affairs inspires us to think that attempts to find expressions
for fusion matrix and boundary correlation functions in the parafermionic
LFT inevitably will require to write  parafermionic generalizations of
$\Gamma_b$ and $S_b$ functions as well. In fact a parafermionic generalization
of $\Gamma_b$ was introduced in \cite{Poghosyan:2016kvd}, where also some
properties of this function were derived.
It is also natural to assume that parafermionic generalization
of $S_b$ function should possess the star-triangle relation as well.

In this paper we would like to connect mentioned topics with the subject which developed over the last
decade in higher dimensional superconformail field theories -- the theory of superconformal indices
\cite{RR2016} described in terms of the elliptic hypergeometric integrals \cite{spi:umnrev}.
So, the standard elliptic gamma function coincides with the superconformal index of chiral superfield
of theories on $S^3\times S^1$ space-time background. Consideration of superconformal indices
of gauge theory on lens space \cite{Benini:2011nc} leads to a particular combination of
elliptic gamma functions with different bases. It was called in \cite{Spiridonov:2016uae}
the rarefied elliptic gamma function due to its special product type representation, using which we
introduce parafermionic hyperbolic gamma function as a particular limit.
 It is built from $S_b$ functions along the same rules by which $\Upsilon_i$ function
 in \cite{Bershtein:2010wz} is built from $\Upsilon_b$ functions.

We show that such parafermionic hyperbolic gamma functions are related
to two computable rarefied hyperbolic
beta integrals, corresponding to two values of a parameter $\epsilon=0,1$.
The one corresponding to $\epsilon=0$ was found earlier in \cite{Gahramanov:2016ilb},
and the second one $\epsilon=1$ is new. Degenerating these hyperbolic beta integrals we obtain
the star-triangle relation for the parafermionic LFT. For the supersymmetric case we compared
obtained results with those derived earlier in \cite{Hadasz:2013bwa} and found that the
star-triangle relation in \cite{Hadasz:2013bwa} in some cases is missing an overall sign.
Thus, it appears that $4d$ superconformal indices
contain a lot of important information about $2d$ systems --
the $2d$ conformal field theories discussed above and integrable
$2d$ lattice spin systems, for which they describe partition
functions \cite{Spiridonov:2010em}. Moreover, it is known that the same
hyperbolic limit of these $4d$ indices describes partition functions of
$3d$ supersymmetric models on the squashed sphere $S^3_b$
\cite{RR2016}.
The present work can be considered as a complement to \cite{DSV}, where
the transition from $4d$ theories to $3d$ ones was reached
by degenerating elliptic hypergeometric integrals to hyperbolic integrals,
-- we add to such a connection a relation to the parafermionic LFT.

We would like to add that, in view of the AGT relation between para-Liouville theory  and superconformal gauge theories on $\mathbb{C}^2/\mathbb{Z}_r$, see e.g. \cite{Alfimov:2011ju,Nishioka:2011jk,Bonelli:2012ny} and references therein, it could be expected that parafermionic hyperbolic gamma functions should arise from the rarefied (or lens) elliptic gamma function.

The paper is organized in the following way. In section 2 we review the necessary formulas on elliptic
gamma functions and the rarefied elliptic beta integral. In section 3 we consider parafermionic hyperbolic gamma functions.
In section 4 we derived a hyperbolic beta integral and star-triangle relation for parafermionic hyperbolic gamma functions. In section 5 we consider in detail the star-triangle relation for supersymmetric case, compare
it with a version of this formula obtained earlier in \cite{Hadasz:2013bwa} and indicate
a sign difference in them.

\section{A rarefied elliptic beta integral}

The standard elliptic gamma function $\Gamma(z;p,q)$ can be defined as an infinite product:
\be
\Gamma(z;p,q)=\prod_{j,k=0}^{\infty}{1-z^{-1}p^{j+1}q^{k+1}\over 1-zp^jq^k}\, ,
\quad |p|,|q|<1\, ,\quad z\in \mathbb{C}^*.
\ee
The lens space elliptic gamma function is defined as a product of two standard elliptic gamma functions with different bases  \cite{Benini:2011nc}.
\bea\label{lensf}
&&\gamma_e(z,m; p,q)=\Gamma(zp^m; p^r,pq)\Gamma(zq^{r-m};q^r,pq)\\ \nonumber
&&=\prod^{\infty}_{j,k=0}{1-z^{-1}p^{-m}(pq)^{j+1}p^{r(k+1)}\over 1-zp^m(pq)^jp^{rk}}
{1-z^{-1}q^m(pq)^{j+1}q^{rk}\over 1-zq^{r-m}(pq)^jq^{rk}}\, ,
\quad m\in\mathbb{Z}.
\eea
As shown in  \cite{Spiridonov:2016uae}, the function (\ref{lensf}) can be written as a special
product of the standard elliptic gamma functions
with bases $p^r$ and $q^r$. For $0\leq m\leq r$ it has the form:
\be
\gamma_e(z,m; p,q)=\prod_{k=0}^{m-1}\Gamma(q^{r-m}z(pq)^k;p^r,q^r)\prod_{k=0}^{r-m-1}\Gamma(p^mz(pq)^k;p^r,q^r)\, ,
\ee
for $m<0$
\be
\gamma_e(z,m; p,q)={\prod_{k=0}^{r-m-1}\Gamma(p^mz(pq)^k;p^r,q^r)
\over \prod_{k=1}^{-m}\Gamma(q^{r-m}z(pq)^{-k};p^r,q^r)}\, ,
\ee
and for $m>r$
\be
\gamma_e(z,m; p,q)={\prod_{k=0}^{m-1}\Gamma(q^{r-m}z(pq)^k;p^r,q^r)\over \prod_{k=1}^{m-r}\Gamma(p^mz(pq)^{-k};p^r,q^r)}\, .
\ee
A convenient normalization of this function was introduced in  \cite{Spiridonov:2016uae}
\bea
&&\Gamma^{(r)}(z,m; p,q) =
(-z)^{m(m-1)\over 2}p^{m(m-1)(m-2)\over 6}q^{-{m(m-1)(m+1)\over 6}}\gamma_e(z,m;p,q),
\eea
which yields $\Gamma^{(1)}(z,m;p,q)=\Gamma(z;p,q)$.
It is this object that was called the rarefied elliptic gamma function.

Let us define a particular combination of such functions
\be
\Delta^{(r)}_e(z,m;t_a,n_a|p,q)={\prod_{a=1}^6\Gamma^{(r)}(t_az,n_a+m+\epsilon; p,q)\Gamma^{(r)}(t_az^{-1},n_a-m; p,q)\over \Gamma^{(r)}(z^2,2m+\epsilon; p,q)\Gamma^{(r)}(z^{-2},-(2m+\epsilon); p,q)},
\ee
\be
\Psi^{(r)}_e(z,m;t_a,n_a|p,q)={\prod_{a=1}^6\gamma_{e}(t_az,n_a+m+\epsilon; p,q)\gamma_{e}(t_az^{-1},n_a-m; p,q)\over \gamma_{e}(z^2,2m+\epsilon; p,q)\gamma_{e}(z^{-2},-(2m+\epsilon); p,q)}.
\ee

It is shown in \cite{Spiridonov:2016uae} that if parameters $t_a$, $n_a$
satisfy the constraints $|t_a|<1$ and the balancing condition
\be
 \prod_{a=1}^6t_a=pq\, ,\quad \sum_{a=1}^6n_a=-3\epsilon\; ,\quad \epsilon=0,1,
\ee
then one has the following integral identity
\be
\kappa^{(r)}\sum_{m=0}^{r-1}\int_{\mathbb{T}}\Delta^{(r)}_e(z,m;t_a,n_a|p,q){dz\over z}=\prod_{1\leq a < b \leq 6}\Gamma^{(r)}(t_at_b,n_a+n_b+\epsilon; p,q),
\ee
where $\mathbb{T}$ is the unit circle of positive orientation and
\be
\kappa^{(r)}={(p^r;p^r)_{\infty}(q^r;q^r)_{\infty}\over 4\pi i}.
\ee
Equivalently, it can be written as
\bea\label{inteiden}
&&\kappa^{(r)}\sum_{m=0}^{r-1}\left({q\over p}\right)^{m^2+m\epsilon}\int_{\mathbb{T}}z^{-\epsilon-2m}\Psi^{(r)}_e(z,m;t_a,n_a|p,q){dz\over z}
\\ \nonumber && \makebox[1em]{}
=\left({q\over p}\right)^{{1\over 2}\sum_{a=1}^6 n_a^2}p^{\epsilon/2}q^{-3\epsilon/2}
(-1)^{\epsilon}\prod_{a=1}^6t_a^{-n_a}\prod_{1\leq a < b \leq 6}\gamma_{e}(t_at_b,n_a+n_b+\epsilon; p,q).
\eea
For $\epsilon=0$ this relation was established first by Kels \cite{Kels}
using a different normalization of the $\gamma_{e}$-functions. Note that
values of the integer parameter $\epsilon$ were reduced to
$0$ and $1$ by admissible shifts of $n_a$.
For $r=1$ one obtains the standard elliptic beta integral \cite{spi:umnrev}.

\section{Parafermionic hyperbolic gamma function}

The function $\Gamma(z;p,q)$ has the following limiting behaviour \cite{ruij}:
\be
\Gamma(e^{-2\pi vy};e^{-2\pi v\omega_1},e^{-2\pi v\omega_2})
\stackreb{=}{ v\to 0}
e^{-\pi(2y-\omega_1-\omega_2)/12v\omega_1\omega_2}\gamma^{(2)}(y;\omega_1,\omega_2),
\ee
where $\gamma^{(2)}(y;\omega_1,\omega_2)$ is the hyperbolic gamma function.
The parameter $v$ approaches to $0$ along the positive real axis
and parameters $\omega_1$ and $\omega_2$ have positive real parts:
 ${\rm Re}(\omega_1)>0$ and ${\rm Re}(\omega_2)>0$.

The function $\gamma^{(2)}(y;\omega_1,\omega_2)$ has the integral representation
\be
\gamma^{(2)}(y;\omega_1,\omega_2)=\exp\left(-\int_0^{\infty}\left({\sinh(2y-\omega_1-\omega_2)x\over 2\sinh(\omega_1x)
\sinh(\omega_2x)}-{2y-\omega_1-\omega_2\over 2\omega_1\omega_2x}\right)\right){dx\over x}\, ,
\ee
and obeys the equations:
\be\label{hp1}
{\gamma^{(2)}(y+\omega_1;\omega_1,\omega_2)\over \gamma^{(2)}(y;\omega_1,\omega_2)}=2\sin{\pi y\over \omega_2}\,  ,\quad
{\gamma^{(2)}(y+\omega_2;\omega_1,\omega_2)\over \gamma^{(2)}(y;\omega_1,\omega_2)}=2\sin{\pi y\over \omega_1}.
\ee
Setting
\be\label{param}
z=e^{-2\pi vy/r}\, ,\quad p=e^{-2\pi v\omega_1/r}\, ,\quad
q=e^{-2\pi v\omega_2/r}\, ,
\ee
one can write
$$
q^{r-m}z(pq)^k=e^{-2\pi  v\left[{y\over r}+\omega_2\left(1-{m\over r}\right)+(\omega_1+\omega_2){k\over r}\right]}\, ,
$$
$$
p^{m}z(pq)^k=e^{-2\pi  v\left[{y\over r}+{m\over r}\omega_1+(\omega_1+\omega_2){k\over r}\right]}\; .
$$
Now one can show that:
\be\label{limg}
\gamma_e\left(e^{-{2\pi vy\over r}}, m;e^{-{2\pi v\omega_1\over r}},
e^{-{2\pi v\omega_2\over r}}\right)\stackreb{=}{ v\to 0}e^{-\pi(2y-\omega_1-\omega_2)/12v\omega_1\omega_2}\Lambda(y,m;\omega_1,\omega_2)\, ,
\ee
where the function $\Lambda(y,m;\omega_1,\omega_2)$ is defined as follows.  For $0\leq m\leq r$ one has
\bea
&&\Lambda(y, m;\omega_1,\omega_2)=\prod_{k=0}^{m-1}\gamma^{(2)}
\left({y\over r}+\omega_2\left(1-{m\over r}\right)+(\omega_1+\omega_2){k\over r};\omega_1,\omega_2\right)
\\ \nonumber && \times
\prod_{k=0}^{r-m-1}\gamma^{(2)}
\left({y\over r}+{m\over r}\omega_1+(\omega_1+\omega_2){k\over r};\omega_1,\omega_2\right)\, ,
\eea
for $m<0$
\bea
\Lambda(y,m;\omega_1,\omega_2)=
{\prod_{k=0}^{r-m-1}\gamma^{(2)}
\left({y\over r}+{m\over r}\omega_1+(\omega_1+\omega_2){k\over r};\omega_1,\omega_2\right)
\over \prod_{k=1}^{-m}\gamma^{(2)}
\left({y\over r}+\omega_2\left(1-{m\over r}\right)
-(\omega_1+\omega_2){k\over r};\omega_1,\omega_2\right)}\, ,
\eea
and for $m>r$
\bea
\Lambda(y,m;\omega_1,\omega_2)={\prod_{k=0}^{m-1}\gamma^{(2)}
\left({y\over r}+\omega_2\left(1-{m\over r}\right)+(\omega_1+\omega_2){k\over r};\omega_1,\omega_2\right)\over \prod_{k=1}^{m-r}\gamma^{(2)}
\left({y\over r}+{m\over r}\omega_1-(\omega_1+\omega_2){k\over r};\omega_1,\omega_2\right)}\, .
\eea
In fact it is enough to consider functions $\Lambda(y,m;\omega_1,\omega_2)$ only for $0\leq m\leq r$.
Recall the quasiperiodicity property  \cite{Spiridonov:2016uae}:
\be
{\gamma_e(z,m+kr;p,q)\over \gamma_e(z,m;p,q)}=\left(-{\sqrt{pq}\over z}\right)^{mk+r{k(k-1)\over 2}}
\left({q\over p}\right)^{k\left({m^2\over 2}+mr{k-1\over 2}+r^2{(k-1)(2k-1)\over 12}\right)},\quad k\in \mathbb{Z}\, .
\ee
In the limit (\ref{limg}) it implies
\be\label{lambdasign}
\Lambda(y,m+kr;\omega_1,\omega_2)=(-1)^{mk+r{k(k-1)\over 2}}\Lambda(y,m;\omega_1,\omega_2)
\ee

Let us study the function $\Lambda(y,m;\omega_1,\omega_2)$ for
the particular choice $r=2$. Eq. (\ref{lambdasign}) implies that
in this case we have only two functions corresponding to $m=0,1$.
For $m=0$ we have:
\be
\Lambda(y,0;\omega_1,\omega_2)=\gamma^{(2)}\left({y\over 2};
\omega_1,\omega_2\right)\gamma^{(2)}\left({y\over 2}
+{\omega_1+\omega_2\over 2};\omega_1,\omega_2\right)\, ,
\ee
and for $m=1$
\be
\Lambda(y,1;\omega_1,\omega_2)=\gamma^{(2)}\left({y\over 2}
+{\omega_2\over 2};\omega_1,\omega_2\right)\gamma^{(2)}
\left({y\over 2}+{\omega_1\over 2};\omega_1,\omega_2\right)\, .
\ee

Setting $\omega_2=b$ and $\omega_1={1\over b}$ and $Q=b+{1\over b}$
and using the notation accepted in conformal field theory literature
\be\label{gams}
\gamma^{(2)}(z;b,1/b)=S_b(z),
\ee
we obtain that
\be\label{l1s}
\Lambda(y,0;b^{-1},b)=S_b\left({y\over 2}\right)
S_b\left({y\over 2}+{Q\over 2}\right)\equiv S_{\rm NS}(y)\equiv S_{1}(y),
\ee
\be\label{l2s}
\Lambda(y,1;b^{-1},b)=S_b\left({y\over 2}+{b\over 2}\right)
S_b\left({y\over 2}+{b^{-1}\over 2}\right)\equiv S_{\rm R}(y)\equiv S_{0}(y).
\ee
The functions $S_{\rm NS}(y)$ and $S_{\rm R}(y)$ appear in numerous aspects of $N=1$ supersymmetric  Liouville conformal field theory.
Subscripts NS and R refer to the Neveu-Schwarz and Ramond sectors respectively.
First defined in  \cite{Fukuda:2002bv} for calculation of  the boundary two-point functions, they played important role
in writing down  fusion and braiding matrices of conformal blocks \cite{Hadasz:2007wi}. It was suggested in \cite{Hadasz:2013bwa}
to denote them as $S_{1}(y)$ and $S_{0}(y)$, respectively, to write  in compact way the corresponding star-triangle relation.

Consider now the functions $\Lambda(y;m;\omega_1,\omega_2)$ for arbitrary $r$:
\bea\nonumber
&&\Lambda(y,m;b^{-1},b)=\prod_{k=0}^{m-1}S_b
\left({y\over r}+b\left(1-{m\over r}\right)+Q{k\over r}\right)
\\  && \makebox[2em]{} \times
\prod_{k=0}^{r-m-1}S_b
\left({y\over r}+{m\over r}b^{-1}+Q{k\over r}\right).
\label{lamom}\eea

Compare them with the  $\Upsilon^{(r)}_m(y)$ functions
defined in \cite{Bershtein:2010wz} for the purpose of
calculation of three-point functions in the parafermionic
Liouville field theory:
\be\label{parfu}
\Upsilon^{(r)}_m(y)=\prod_{j=1}^{r-m}\Upsilon_b\left({y+mb^{-1}+(j-1)Q\over r}\right)
\prod_{j=r-m+1}^{r}\Upsilon_b\left({y+(m-r)b^{-1}+(j-1)Q\over r}\right)\, .
\ee
Let us replace $\Upsilon_b$ by $S_b$ in expression (\ref{parfu}).
Then the substitution $j=k+1$ in its first product yields precisely
the second product in (\ref{lamom}). Similarly, the substitution $j=k+r-m+1$
converts its second product to the first one in (\ref{lamom})
because $b^{-1}-Q=-b$. So, we have intriguing exact structural correspondence
between the functions (\ref{parfu}) and (\ref{lamom}).

For this reason we call $\Lambda(y;m;\omega_1,\omega_2)$
the parafermionic hyperbolic gamma function. It should play the
same role in the construction of parafermionic fusion matrices
as $\Upsilon^{(r)}_m(y)$ serves the correlation functions.
Applying the limit (\ref{limg}) to expression (\ref{lensf}) one can
derive another expression for it
\be\label{gk}
\Lambda(y,m;\omega_1,\omega_2)=\gamma^{(2)}\left({y+m\omega_1\over r};\omega_1,{\omega_1+\omega_2\over r}\right)
\gamma^{(2)}\left({y+(r-m)\omega_2\over r};\omega_2,{\omega_1+\omega_2\over r}\right),
\ee
which was obtained in \cite{Gahramanov:2016ilb,IY,Nieri:2015yia}.
Using equations (\ref{hp1}) one can easily show that (\ref{gk})
satisfies (\ref{lambdasign}).

\section{Integral identities for parafermionic hyperbolic gamma functions}

Now we apply the limit (\ref{limg}) to the rarefied elliptic beta integral evaluation (\ref{inteiden}).
For that we set additionally to (\ref{param}) the parameterization:
\be
t_a=e^{-{2\pi v s_a\over r}}\, ,\quad \sum_{a=1}^6 s_a=\omega_1+\omega_2
\ee
and take the limit $v\to 0^+$. As a result, we obtain the following identity
representing a rarefied hyperbolic beta integral evaluation
\bea\nonumber
&&\int_{-i\infty}^{i\infty}\sum_{m=0}^{r-1}{\prod_{a=1}^6\Lambda(y+s_a,n_a+m+\epsilon;\omega_1,\omega_2)
\Lambda(-y+s_a,n_a-m;\omega_1,\omega_2)
\over \Lambda(2y,2m+\epsilon;\omega_1,\omega_2)\Lambda(-2y,-
(2m+\epsilon);\omega_1,\omega_2)}{dy\over i\sqrt{\omega_1\omega_2}}
\\  &&  \makebox[2em]{}
= 2r(-1)^{\epsilon}\prod_{1\leq a < b \leq 6}
\Lambda(s_a+s_b,n_a+n_b+\epsilon;\omega_1,\omega_2).
\label{newidz}\eea

For $\epsilon=0$ this evaluation was derived in \cite{Gahramanov:2016ilb} and
for $\epsilon=1$ it is a new result. Reductions of the ordinary $r=1$ elliptic
hypergeometric integrals to the hyperbolic level are systematically considered
in \cite{BultPhD}. They are based on a rigorous justification for such
transitions established in \cite{rai:limits}.

To derive from (\ref{newidz}) the parafermionic star-triangle relation
we should elaborate asymptotic properties of the $\Lambda(y,m;\omega_1,\omega_2)$
function.

The function $\gamma^{(2)}(y;\omega_1,\omega_2)$ has the following
asymptotics \cite{Kharchev:2001rs}:
\be
\stackreb{\lim}{y\to \infty}e^{{i\pi\over 2}B_{2,2}(y;\omega_1,\omega_2)}\gamma^{(2)}(y;\omega_1,\omega_2)=1,
\quad {\rm for}\; {\rm arg}\;\omega_1<{\rm arg}\; y<{\rm arg}\;\omega_2+\pi,
\ee
\be
\stackreb{\lim}{y\to \infty}e^{-{i\pi\over 2}B_{2,2}(y;\omega_1,\omega_2)}\gamma^{(2)}(y;\omega_1,\omega_2)=1,
\quad {\rm for}\; {\rm arg}\;\omega_1-\pi<{\rm arg}\; y<{\rm arg}\;\omega_2,
\ee
where $B_{2,2}(y;\omega_1,\omega_2)$  is the second order Bernoulli polynomial:
\be
B_{2,2}(y;\omega_1,\omega_2)={y^2\over \omega_1\omega_2}
-{y\over \omega_1}-{y\over \omega_2}+{1\over 6}
\left({\omega_1\over \omega_2}+{\omega_2\over \omega_1}\right)+{1\over 2}.
\ee
Because of (\ref{gk}) this implies that $\Lambda(y,m,\omega_1,\omega_2)$ function
has similar asymptotics with $B_{2,2}(y;\omega_1,\omega_2)$ replaced by:
\bea\nonumber
&& B_{2,2}\left({y+m\omega_1\over r}; \omega_1, {\omega_1+\omega_2\over r}\right)+
B_{2,2}\left({y+(r-m)\omega_2\over r}; \omega_2, {\omega_1+\omega_2\over r}\right)
\\ \nonumber && \makebox[2em]{}
={y^2\over \omega_1\omega_2r}-{y\over r\omega_1}-{y\over r\omega_2}
+{1\over 6r}\left({\omega_1\over \omega_2}+{\omega_2\over \omega_1}\right)
+{m^2\over r}-m+{r\over 6}+{1\over 3r}
\\ && \makebox[2em]{}
= {1\over r}B_{2,2}(y;\omega_1,\omega_2)+{m^2\over r}-m+{r\over 6}-{1\over 6r}.
\label{asympber}\eea

Let us reparameterize $s_a$ in (\ref{newidz}) in the following asymmetric way
\bea
&&s_a= f_a+i\mu, \quad a=1,2,3, \quad
s_{a+3}= g_{a}-i\mu, \quad a=1,2,3,
\eea
which preserves the balancing condition. We denote also
\be
n_{a+3}\equiv l_a,\quad a=1,2,3.
\ee
So, we have
\be
\sum_{a=1}^3(f_a+g_a)=\omega_1+\omega_2
\ee
and
\be
\sum_{a=1}^3(n_a+l_a)=-3\epsilon.
\ee

Now we shift the integration variable $y\to y-i\mu$ and take the limit
$\mu\to +\infty$ using the asymptotics of $\Lambda(y,m,\omega_1,\omega_2)$.
Since the integrand is an even function (in fact the parity
transformation reshuffles the terms keeping the sum intact), one can write:
\bea \nonumber
&&
2\int_{0}^{i\infty}\sum_{m=0}^{r-1}\left[{\prod_{a=1}^3\Lambda(y+f_a+i\mu,n_a+m+\epsilon;\omega_1,\omega_2)
\Lambda(y+g_a-i\mu,l_{a}+m+\epsilon;\omega_1,\omega_2)
\over \Lambda(2y,2m+\epsilon;\omega_1,\omega_2)\Lambda(-2y,-
(2m+\epsilon);\omega_1,\omega_2)}\right.
\\&&
\times \left.\prod_{a=1}^3\Lambda(-y+f_a+i\mu,n_a-m;\omega_1,\omega_2)\Lambda(-y+g_a-i\mu,l_{a}-m;\omega_1,\omega_2)\right]
{dy\over i\sqrt{\omega_1\omega_2}}
\\ \nonumber &&
=2\int_{-i\mu}^{i\infty}\sum_{m=0}^{r-1}\prod_{a=1}^3\Lambda(y+f_a,n_a+m+\epsilon;\omega_1,\omega_2)
\Lambda(-y+g_a,l_a-m;\omega_1,\omega_2)e^{{i\pi\over 2}\sigma_1}
{dy\over i\sqrt{\omega_1\omega_2}},
\eea
where in the limit $\mu\to \infty$
\bea\label{sigma1} &&\makebox[-2em]{}
\sigma_1={1\over r}\sum_{a=1}^3\left[B_{2,2}(y+g_a-2i\mu)
-B_{2,2}(-y+f_a+2i\mu)\right]
\\ \nonumber &&
-{1\over r}B_{2,2}(2y-2i\mu)+ {1\over r}B_{2,2}(-2y+2i\mu)
\\
&&+\sum_{a=1}^3 \left[{1\over r}(l_{a}+m+\epsilon)^2-(l_{a}+m+\epsilon)-{1\over r}(n_{a}-m)^2+(n_{a}-m)\right]
+4m+2\epsilon.
\nonumber\eea

On the right-hand side we have
\be
2(-1)^{\epsilon}r\prod_{a,b=1}^3\Lambda(f_a+g_b,n_a+l_b+\epsilon;\omega_1,\omega_2)
e^{{i\pi\over 2}\sigma_2},
\ee
where
\bea\label{sigma2}
&&\sigma_2={1\over r}\sum_{1\leq a<b\leq 3}\left[B_{2,2}(g_a+g_b-2i\mu)
-B_{2,2}(f_a+f_b+2i\mu)\right]
\\ \nonumber &&
+\sum_{1\leq a<b\leq 3}\left[{1\over r}(l_a+l_b+\epsilon)^2
-(l_a+l_b+\epsilon)-{1\over r}(n_a+n_b+\epsilon)^2+(n_a+n_b+\epsilon)\right].
\eea

Similar to the considerations of \cite{BultPhD,Spiridonov:2010em} for $r=1$ case,
it can be checked that all $B_{2,2}$-terms appearing in (\ref{sigma1})
and (\ref{sigma2}) cancel each other.
Taking care about the rest yields:
\bea\nonumber  \makebox[-2em]{}
&&\int_{-i\infty}^{i\infty}\sum_{m=0}^{r-1}(-1)^m\prod_{a=1}^3\Lambda(y+f_a,n_a+m+\epsilon;\omega_1,\omega_2)
\Lambda(-y+g_a,l_a-m;\omega_1,\omega_2)
{dy\over i\sqrt{\omega_1\omega_2}}
\\ &&  \makebox[2em]{}
=(-1)^{\epsilon+n_1+n_2+n_3}r\prod_{a,b=1}^3
\Lambda(f_a+g_b,n_a+l_b+\epsilon;\omega_1,\omega_2).
\label{namer}\eea
This is the desired star-triangle relation for the parafermionic hyperbolic
gamma  functions.
Note that using (\ref{lambdasign}) we can always bring all the functions
to the basic domain $0\leq m\leq r$. For $r=1$ one gets the star-triangle relation for the Faddeev-Volkov model
\cite{Bazhanov:2007vg,volkov}.

\section{Supersymmetric Liouville model case}

In this section we study in detail relation (\ref{namer}) for supersymmetric
hyperbolic gamma functions, which correspond to the choice $r=2$.
So, we set in  (\ref{namer})  $r=2$ and  $\omega_1=1/b$, $\omega_2=b$.
Also we represent integer variables $n_a$ and $l_a$ in the form
\bea
&& n_a=2k_a+\nu_a\, ,\quad \nu_a=0,1\, ,\quad a=1,2,3,\\ \nonumber
&& l_a=2h_a+\mu_a\, ,\quad \mu_a=0,1\, ,\quad a=1,2,3,
\eea
for some integers $k_a$ and $h_a$.

Consider first the case $\epsilon=1$.
Using (\ref{l1s}), (\ref{l2s}) and (\ref{lambdasign}), we can write:
\be\label{1fa}
\Lambda(y+f_a,n_a+m+1;b^{-1},b)=(-1)^{k_a(\nu_a+m+1)}
(-1)^{\nu_a m}S_{\nu_a+m}(y+f_a),
\ee
\be\label{1gb}
\Lambda(-y+g_a,l_a-m;b^{-1},b)=(-1)^{h_a(\mu_a-m)}
(-1)^{(\mu_a+1) m}S_{\mu_a+m+1}(-y+g_a),
\ee
\be\label{1fagb}
\Lambda(f_a+g_b,n_a+l_b+1;b^{-1},b)=(-1)^{(k_a+h_b)(\nu_a
+\mu_b+1)}(-1)^{\nu_a\mu_b}S_{\nu_a+\mu_b}(f_a+g_b).
\ee
The subscript $a$ in $S_a(y)$ is defined mod 2: $S_{a+2k}(y)\equiv S_a(y)$.

Inserting (\ref{1fa})-(\ref{1fagb}) in (\ref{namer}) we obtain
\bea \nonumber
&&\sum_{m =0,1}(-1)^{m (1+\sum_a (\nu_a+\mu_a))/2}
\int {dx\over i}\prod_{a=1}^3 S_{m+\nu_a}(x+f_a)S_{1+m+\mu_a}(-x+g_a)
\\ && \makebox[2em]{}
=2(-1)^{\left(\sum\mu_a\right) (1+\sum_a (\nu_a+\mu_a))/2}
\prod_{a,b=1}^3 S_{\nu_a+\mu_b}(f_a+g_b)\, ,
\label{sinteg}\eea
\be\label{vmn2}
\sum_a(\nu_a+\mu_a)=1\; {\rm mod}\; 2\, ,
\ee
and
\be
\sum_a(f_a+g_a)=Q\, .
\ee

Consider now the case $\epsilon=0$.
Using (\ref{l1s}), (\ref{l2s}) and (\ref{lambdasign}) we can write:
\be\label{1fa'}
\Lambda(y+f_a,n_a+m;b^{-1},b)=(-1)^{k_a(\nu_a+m)}S_{\nu_a+m+1}(y+f_a),
\ee
\be\label{1gb'}
\Lambda(-y+g_a,l_a-m;b^{-1},b)=(-1)^{h_a(\mu_a-m)}
(-1)^{(\mu_a+1) m}S_{\mu_a+m+1}(-y+g_a),
\ee
\be\label{1fagb'}
\Lambda(f_a+g_b,n_a+l_b;b^{-1},b)=(-1)^{(k_a+h_b)(\nu_a+\mu_b)}
S_{\nu_a+\mu_b+1}(f_a+g_b).
\ee

Inserting (\ref{1fa'})-(\ref{1fagb'}) in (\ref{namer}) we obtain
\bea\nonumber
&&
\sum_{m =0,1}(-1)^{m (\sum_a (\mu_a-\nu_a))/2}\int {dx\over i}
\prod_{a=1}^3 S_{m+\nu_a+1}(x+f_a)
S_{1+m+\mu_a}(-x+g_a)
\\  && \makebox[2em]{}
=2(-1)^{\left(\sum\mu_a\right) (\sum_a (\mu_a-\nu_a))/2}
\prod_{a,b=1}^3 S_{\nu_a+\mu_b+1}(f_a+g_b)\, ,
\label{sinteg2}\eea
\be\label{vmn22}
\sum_a(\nu_a+\mu_a)=0\; {\rm mod}\; 2\, ,
\ee
and
\be
\sum_a(f_a+g_a)=Q\, .
\ee
It is obvious that (\ref{sinteg}), (\ref{vmn2}) and (\ref{sinteg2}),
(\ref{vmn22}) are related by the transformation $\nu_a\to 1-\nu_a,\, a=1,2,3$,
i.e. we have only one independent relation.

Comparing (\ref{sinteg}) with the star-triangle relation found in
\cite{Hadasz:2013bwa}, we see that they coincide in all aspects
besides of the overall sign in the right-hand side
$(-1)^{\left(\sum\mu_a\right) (1+\sum_a (\nu_a+\mu_a))/2}$ present
in our formula. We suggest the following independent check of the presence
of this multiplier in a particular case, when it is equal to $-1$.
Such a situation takes place only when both $\sum\mu_a$ and
${(1+\sum_a (\nu_a+\mu_a))\over 2}$ are odd.
The following choice of the parameters obviously satisfies both conditions:
\be
\nu_1=0,\quad \nu_2=0,\quad \nu_3=0
\ee
and
\be
\mu_1=1,\quad \mu_2=0, \quad \mu_3=0\, .
\ee

Substituting these values in (\ref{sinteg}) we obtain:
\bea\nonumber
&&
\int {dx\over i}\left[S_0(x+f_1)S_0(x+f_2)S_0(x+f_3)
S_0(-x+g_1)S_1(-x+g_2)S_1(-x+g_3)\right.
\\ \nonumber && \makebox[1em]{}
-\left.S_1(x+f_1)S_1(x+f_2)S_1(x+f_3)S_1(-x+g_1)S_0(-x+g_2)S_0(-x+g_3)\right]
\\ \nonumber && \makebox[1em]{}
=-2S_1(f_1+g_1)S_0(f_1+g_2)S_0(f_1+g_3)
 \\  && \makebox[1em]{}
\times S_1(f_2+g_1)S_0(f_2+g_2)S_0(f_2+g_3)S_1(f_3+g_1)S_0(f_3+g_2)S_0(f_3+g_3).\quad
\label{starsuper}\eea

Let us study this integral directly in the limit $f_1+g_1\to 0$
 and compare it with the suggested right-hand side expression.

As a warm-up exercise consider at the beginning this question for
the ``bosonic'' star-triangle identity
\bea\label{bst}
\int {dx\over i}\prod_{j=1}^3S_b(x+f_j)S_b(-x+g_j)
=\prod_{j,k=1}^3 S_b(f_j+g_k).
\eea
Recall that the function $S_b(x)$ is meromorphic with poles at $x=-nb-mb^{-1}$, and  zeros at
$x=Q+nb+mb^{-1}$, where $n$ and $m$ are non-negative integers.
Around zero $x=0$ the $S_b(x)$ function has the behavior:
\be\label{sbres}
\stackreb{\rm lim}{x\to 0}xS_b(x)= {1\over 2\pi }.
\ee

Take the limit $f_1+g_1\to 0$  in a way that $-f_1$ and $g_1$ approach
to a point $A$ of imaginary axis ($A\in i\mathbb{R}$) from different sides.
Without loss of generality we can assume that $-f_1$ moves to
this point from the left side and $g_1$ comes from the right side.
This results in the pinching of the integration contour (the imaginary
axis) by two poles.
Consider the left-hand side integral as a function of parameters $f_i$ and $g_i$.
Let us show  that pinching of the contour results in the pole singularity
of this function $1/(f_1+g_1)$ and compute its leading asymptotics.
For that we deform the integration contour and force it to cross over the
point $x=-f_1$ and pick up the corresponding pole residue determined by
the integral over small circle around $-f_1$. The integral
over deformed contour is finite and the singularity can arise only from the
taken residue. According to (\ref{sbres}) the integrand around the point
$x=-f_1\approx g_1$  takes the asymptotic form:
\be
{1\over 4i\pi^2 (x+f_1)(-x+g_1)}S_b(x+f_2)S_b(x+f_3)S_b(-x+g_2)S_b(-x+g_3).
\ee
Then, by the Cauchy theorem the integral over small circle around
this point is equal to
\be
{1\over 2\pi (f_1+g_1)}S_b(-f_1+f_2)S_b(-f_1+f_3)S_b(f_1+g_2)S_b(f_1+g_3).
\ee

On the other hand, we see that the right-hand side expression
in (\ref{bst}) indeed has
the pole singularity at $f_1+g_1\to 0$ coming from the $j=k=1$ multiplier.
The rest can be seen to yield the same result due to the balancing
condition, which in this limit takes the form
$f_2+f_3+g_2+g_3=Q$,  and relation $S_b(x)S_b(Q-x)=1$.
The same situation will take place if we take the limit $f_1+g_1\to 0$
in an asymmetric way, i.e. for an arbitrary eventual value of $f_1$.
For instance, we may deform the integration contour close
to a fixed point $-f_1$ and in the limit $g_1\to -f_1$ we come inevitably
to pinching of the contour which leads to the same singular
asymptotics for the integral.

Now let us get back to the integral (\ref{starsuper}).
First let us indicate necessary properties of the functions $S_0(x)$ and $S_1(x)$.
The function $S_0(x)$ has zeros at $x=Q+nb+mb^{-1}$ and poles
at $x=-mb-nb^{-1}$, where $m$ and $n$ are both non-negative integers
and $m+n$ is odd.
The function $S_1(x)$ has zeros at $x=Q+nb+mb^{-1}$ and poles
at $x=-mb-nb^{-1}$, where $m$ and $n$ are both non-negative integers
and $m+n$ is even.
The function $S_1(x)$ near zero has the behavior:
\be\label{s1pol}
\stackreb{\rm lim}{x\to 0}xS_1(x)={1\over \pi }.
\ee
Also we have
\be\label{refl10}
S_0(x)S_0(Q-x)=1, \quad S_1(x)S_1(Q-x)=1.
\ee
In the same limit $f_1+g_1\to 0$ the poles of $S_0(x)$ functions
in (\ref{starsuper}) do not pinch the contour ($S_0(0)$ is regular)
and the contribution from the first term in the integrand
remains finite. The pole singularity is produced
only by the second term in the integrand.
Using (\ref{s1pol}) one can see that around the point $x=-f_1$
the integrand asymptotically takes the form
\be\label{ab12'}
-{1\over i\pi^2(x+f_1)(-x+g_1)}S_1(x+f_2)S_1(x+f_3)S_0(-x+g_2)S_0(-x+g_3).
\ee
Again, by the Cauchy theorem the integral over the small circle
around $x=-f_1$ is equal to
\be\label{ab12}
-{2\over \pi(f_1+g_1)}S_1(-f_1+f_2)S_1(-f_1+f_3)S_0(f_1+g_2)S_0(f_1+g_3).
\ee
It is easy to see that, due to the balancing condition and properties
(\ref{s1pol}), (\ref{refl10}), the asymptotics of the
right-hand side expression in (\ref{starsuper}) indeed coincides
with (\ref{ab12}) with the correct sign.

\section{Conclusion}

To conclude, in this work we established a link between the superconformal
indices of $4d$ SCFTs on the lens space, the corresponding rarefied
elliptic hypergeometric functions and the parafermionic Liouville model.
The parafermionic star-triangle relation (\ref{namer}) should play a
proper role in the consideration of corresponding LFT fusion matrices.
Following the logic of the present work it would be also
interesting to investigate the hyperbolic degeneration of the rarefied elliptic
hypergeometric function $V^{(r)}$ constructed in \cite{Spiridonov:2016uae}
and search for its proper parafermionic, or supersymmetric for $r=2$
interpretation.

One of the relevant topics which we skipped in the present note concerns
partition functions of supersymmetric $3d$ field theories described by hyperbolic
integrals. Our relations (\ref{newidz}) and (\ref{namer}) should describe
dualities of certain models on
the manifold $S^3/\mathbb{Z}_r$ similar to the $r=1$ cases \cite{RR2016}.
Indeed, in \cite{IY} a number of such dualitites has been investigated,
but coincidence of dual partition functions was established only
numerically. It would be interesting to analyze whether the
corresponding conjectural identities are consequences of (\ref{newidz})
or hyperbolic limits of other identities from \cite{Spiridonov:2016uae}, or they
describe somewhat different systems.

\smallskip

{\bf Acknowledgements.}
This work is partially supported by the Russian Science Foundation (project no. 14-11-00598).

\end{document}